\documentclass{article}
\ifx\pdfoutput\undefined\else\pdfoutput=1\fi

\usepackage[preprint]{neurips_2024}

\usepackage[utf8]{inputenc}
\usepackage[T1]{fontenc}
\usepackage{hyperref}
\usepackage{url}
\usepackage{graphicx}
\usepackage{booktabs}
\usepackage{multirow}
\usepackage{array}
\usepackage{amsmath}
\usepackage{amssymb}
\usepackage{enumitem}
\usepackage{longtable}
\usepackage{tabularx}

\title{Aesthetic Experience and Educational Value in Co-creating Art with Generative AI: \\
Evidence from a Survey of Young Learners}

\author{%
  Chengyuan Zhang\thanks{Equal contribution.}\qquad Suzhe Xu$^*$ \\
  Huaqiao University
}

\begin{document}

\maketitle

\begin{abstract}
This study investigates the aesthetic experience and educational value of collaborative artmaking with generative artificial intelligence (AI) among young learners and art students. Based on a survey of 112 participants, we examine how human creators renegotiate their roles, how conventional notions of originality are challenged, how the creative process is transformed, and how aesthetic judgment is formed in human--AI co-creation. Empirically, participants generally view AI as a partner that stimulates ideation and expands creative boundaries rather than a passive tool, while simultaneously voicing concerns about stylistic homogenization and the erosion of traditional authorship. Theoretically, we synthesize Dewey's aesthetics of experience, Ihde's postphenomenology, and actor--network theory (ANT) into a single analytical framework to unpack the dynamics between human creators and AI as a non-human actant. Findings indicate (i) a fluid subjectivity in which creators shift across multiple stances (director, dialogic partner, discoverer); (ii) an iterative, dialogic workflow (intent--generate--select--refine) that centers critical interpretation; and (iii) an educational value shift from technical skill training toward higher-order competencies such as critical judgment, cross-modal ideation, and reflexivity. We argue that arts education should cultivate a \emph{critical co-creation} stance toward technology, guiding learners to collaborate with AI while preserving human distinctiveness in concept formation, judgment, and meaning-making.
\end{abstract}


\section{Introduction}
Arts education aims to develop perception, imagination, judgment, and humanistic literacy through aesthetic activity. With rapid digitization and the integration of AI into curricula, text-to-image systems (e.g., Midjourney, DALL\'E 2, Stable Diffusion \cite{midjourneyWeb,ramesh2022hierarchical,rombach2022high}) have altered the logic of artistic production by partially decoupling ideation from manual execution, enabling non-experts to realize complex visuals rapidly. This shift raises foundational questions for arts education: What is the role of the human creator when AI can instantly render imagery? How should authorship and originality be reconceived when outcomes are co-produced by humans and machines? And how are aesthetic decisions formed amid stochastic, iterative human--AI interaction?

To address these questions, we conducted a survey titled ``Co-creating Art with Generative AI: Aesthetic Experience and Educational Value'' with 112 valid responses, focusing on young learners (primarily university students) who are both early adopters of digital technologies and actively forming artistic worldviews. We analyze (i) role perception in human--AI collaboration, (ii) views on originality, (iii) transformations in creative process, and (iv) perceived educational value. We then interpret results through Dewey's aesthetics of experience, Ihde's postphenomenology, and actor--network theory (ANT), and conclude with curricular and assessment recommendations for arts education in the age of generative AI.

\section{Related Work and Theoretical Background}
\paragraph{Dewey: Art as Experience.} Dewey situates art in dynamic, unified experience driven by the reciprocity of doing and undergoing \cite{dewey1934art}. In generative AI artmaking, prompt writing is the creator's \emph{doing}, while the system's output is an environmental \emph{undergoing} that often introduces \emph{resistance} (mismatch with intent). Iterative cycles of prompting and interpretation embody Dewey's full aesthetic experience and shift evaluative focus from end-products to process.

\paragraph{Ihde: Postphenomenology of Human--Technology Relations.} Ihde articulates three relations: (i) \emph{embodiment} (technology becomes transparent, extending perception), (ii) \emph{hermeneutic} (technology offers representations to be read), and (iii) \emph{alterity} (technology is encountered as a quasi-other) \cite{ihde1990technology}. Prompt fluency reflects embodiment; reading generated images reflects hermeneutics; surprise from unanticipated outputs indexes alterity. We use these to interpret role fluidity reported by participants.

\paragraph{Actor--Network Theory (ANT).} ANT views agency as distributed across heterogeneous networks of human and non-human actants \cite{latour2005reassembling,callon1986some}. In AI artmaking, creators, models, training data, prompts, and compute platforms jointly shape outcomes. Prompt engineering is a central \emph{translation} that enrolls other actants. From this vantage, originality is not an intrinsic property of a solitary author but an emergent result of networked collaboration.

\section{Methods}
\paragraph{Participants.} We collected 112 valid responses. The sample is predominantly higher-education students (85.71\%), spanning arts-related (34.82\%) and non-arts majors (50.89\%).

\paragraph{Instrument.} The questionnaire comprises demographics, Likert-scale items on role perception, single- and multiple-choice items on originality and creative workflows, and items on educational value. Full wording appears in Appendix~\ref{app:questionnaire}.

\paragraph{Procedure and Analysis.} The survey was administered online. We report descriptive statistics (frequencies, percentages, means, standard deviations), and cross-tabulate select items by background where relevant. Numerical summaries are reproduced below; detailed items and full distributions appear in the appendix.

\section{Results}
\subsection{Participant profile and AI use}
Table~\ref{tab:profile} summarizes participant characteristics and AI usage frequency (N=112).

\begin{table}[t]
\centering
\caption{Participant profile and AI use frequency (N=112).}
\label{tab:profile}
\begin{tabular}{llrr}
\toprule
\textbf{Category} & \textbf{Option} & \textbf{Count} & \textbf{Percent} \\
\midrule
\multirow{3}{*}{Main identity} & Undergraduate/Graduate student & 96 & 85.71\% \\
 & High school student & 4 & 3.57\% \\
 & Independent learner/Other & 12 & 10.72\% \\
\midrule
\multirow{3}{*}{Discipline} & Arts-related major & 39 & 34.82\% \\
 & Non-arts major & 57 & 50.89\% \\
 & Other & 16 & 14.29\% \\
\midrule
\multirow{3}{*}{AI use frequency} & Occasional (a few times/month) & 59 & 52.68\% \\
 & Frequent (a few times/week) & 43 & 38.39\% \\
 & Heavy (almost daily) & 10 & 8.93\% \\
\bottomrule
\end{tabular}
\end{table}

\noindent\textit{Description.} Generative AI is widely known among the target population but not yet a daily practice. A non-trivial fraction are heavy users (8.93\%), suggesting deep integration for a subset of learners.

\subsection{Role perception in human--AI collaboration}
We asked participants to rate role-perception statements (1--5; higher indicates stronger agreement). Table~\ref{tab:role} reports means (M), standard deviations (SD), and the share selecting 4 or 5.

\begin{table}[t]
\centering
\caption{Role perception in collaboration (N=112; 1--5 scale).}
\label{tab:role}
\begin{tabular}{lccc}
\toprule
\textbf{Statement} & \textbf{M} & \textbf{SD} & \textbf{Agree (4/5)} \\
\midrule
I am the director; AI merely executes instructions. & 3.85 & 1.02 & 68.75\% \\
Creative subjectivity is dynamic during the process. & 3.71 & 0.98 & 62.49\% \\
AI outputs often exceed expectations and feed back my creativity. & 4.02 & 0.89 & 62.50\% \\
AI and I are equal co-creation partners. & 3.25 & 1.15 & 41.07\% \\
\bottomrule
\end{tabular}
\end{table}

\noindent\textit{Description.} The highest mean (M=4.02) indicates surprise-driven creativity consistent with Ihde's alterity relation. Strong endorsement of the director stance (68.75\%) reflects embodiment/tool transparency. A majority also recognizes dynamic subjectivity, i.e., fluid shifts across relations.

\subsection{Originality, process, conflict strategies, and educational value}
Key judgments about originality, workflow, conflict handling, and learning benefits are summarized below.

\begin{table}[t]
\centering
\caption{Originality (single choice).}
\label{tab:orig}
\begin{tabular}{lr}
\toprule
\textbf{Option} & \textbf{Percent} \\
\midrule
Yes, originality needs redefinition (e.g., prompt design as authorship). & 53.57\% \\
No, AI is merely a tool; originality remains human. & 35.71\% \\
Uncertain. & 10.72\% \\
\bottomrule
\end{tabular}
\end{table}

\begin{table}[t]
\centering
\caption{Perceived change in creative process (single choice).}
\label{tab:process}
\begin{tabular}{lr}
\toprule
\textbf{Option} & \textbf{Percent} \\
\midrule
Shift to an iterative loop: intent--generate--select--refine. & 46.43\% \\
Still a linear ``ideation--execution'' process. & 21.43\% \\
Other / little change. & 32.14\% \\
\bottomrule
\end{tabular}
\end{table}

\begin{table}[t]
\centering
\caption{First-choice strategy when outputs conflict with intent.}
\label{tab:conflict}
\begin{tabular}{lr}
\toprule
\textbf{Option} & \textbf{Percent} \\
\midrule
Accept AI output and adjust creative direction. & 37.50\% \\
Revise prompts repeatedly until satisfied. & 37.50\% \\
Abandon AI; return to traditional methods. & 9.82\% \\
Other. & 15.18\% \\
\bottomrule
\end{tabular}
\end{table}

\begin{table}[t]
\centering
\caption{Perceived benefits for arts learning (multiple choice).}
\label{tab:benefits}
\begin{tabular}{lr}
\toprule
\textbf{Option} & \textbf{Percent} \\
\midrule
Cultivating critical filtering and judgment. & 71.43\% \\
Stimulating reflection on traditional artistic languages. & 60.71\% \\
Fostering cross-modal creative thinking. & 50.00\% \\
Improving executional skills (e.g., composition). & 35.71\% \\
\bottomrule
\end{tabular}
\end{table}

\noindent\textit{Description.} A majority favors reconceptualizing originality, with process perceived as iterative and dialogic. Conflict strategies split between embracing serendipity and asserting intent through prompt iteration. Learning gains concentrate on higher-order cognition rather than tool-specific techniques.

\section{Analysis and Discussion}
\subsection{Role fluidity: a postphenomenological account}
High endorsement of the director stance aligns with embodiment relations, while the strongest mean concerns AI-as-surprise (alterity). The recognition that subjectivity is dynamic suggests creators traverse embodiment, hermeneutic, and alterity relations within a single creative episode, toggling between efficient execution, interpretive reading of images, and openness to the unexpected. Subjectivity in AI-mediated creation is thus best described as a processual, negotiated becoming rather than a fixed identity.

\subsection{Reconstructing originality: an actor--network perspective}
The preference to redefine originality coheres with ANT's distributed agency. Participants tacitly recognize prompts, models, and training data as consequential actants. When prompt craft is recognized as creative contribution, originality shifts from a property of a solitary originator to an emergent outcome of networked translation and coordination. This reframing invites legal and ethical innovation toward contribution-sensitive attribution rather than single-author essentialism.

\subsection{Dialogic process and the educational shift}
Nearly half of participants describe a looped workflow epitomizing Dewey's doing/undergoing rhythm. Each generation is an undergoing; each prompt revision a renewed doing. The educational center of gravity moves accordingly: from craft execution to meta-creative capacities---critical selection, conceptual synthesis, and meaning-making under uncertainty. In this sense, co-creation itself functions as aesthetic education in \emph{systems literacy} and \emph{reflexive judgment}.

\section{Conclusion and Recommendations}
\paragraph{Conclusions.} (i) Creator roles are hybrid and fluid across human--technology relations; (ii) originality is increasingly viewed as distributed across the human--AI network with prompts as pivotal translators; and (iii) educational value concentrates on higher-order cognition rather than merely executional technique.

\paragraph{Recommendations.}
(1) \emph{Curricular integration}: treat generative AI as core content beyond tool operation, including prompt design, ethics/bias, techno-art histories, and relevant IP topics. (2) \emph{Pedagogical shift}: instructors act as guides and scaffolders in project-based learning that explores the limits of AI and debates human--AI relations and authorship. (3) \emph{Core competencies}: cultivate \emph{critical co-creation}---collaborating effectively with AI while maintaining critical scrutiny of outputs, styles, and biases to consolidate distinctive artistic viewpoints. (4) \emph{Ethics and assessment}: develop co-creation guidelines (e.g., transparency about methods/data) and multi-dimensional assessment that evaluates process evidence and conceptual innovation, not only final artifacts.

\paragraph{Limitations and future work.} The study is limited by its sample size and reliance on self-reports. Future work should triangulate with classroom observations, longitudinal studies, and qualitative interviews to deepen and validate these findings.


\newpage
\appendix
\section{Questionnaire and additional descriptive statistics}
\label{app:questionnaire}
This appendix reproduces the questionnaire items and selected descriptive statistics used in the study. Items are translated from Chinese; numbering follows the deployed instrument.

\subsection*{A. Questionnaire (translated)}
\begin{enumerate}[leftmargin=*]
\item Identity (single choice): Middle school student; University student (arts-related); University student (non-arts); Independent arts learner; Other.
\item Frequency of using AI tools for creation (single choice): Never; Occasional (1--5 times); Monthly; Weekly/dependent.
\item Role perception in AI collaboration (matrix, 1--5): (A) I am the director; AI executes; (B) Equal co-creation partners; (C) AI outputs often exceed expectations and feed back creativity; (D) Creative subjectivity changes dynamically.
\item Has AI collaboration changed your understanding of originality? (single choice): Yes, originality should be redefined (e.g., prompt as creation); No; Uncertain.
\item How are aesthetic decisions formed? (multiple choice): (A) I set style/theme, AI executes; (B) Iterative prompt adjustment and selection; (C) AI sparks new ideas; (D) Rely on default styles; (E) Other.
\item How do you negotiate creative conflicts? (ranking): Accept AI's unconventional output; Revise prompt repeatedly; Fuse multiple AI results; Abandon AI; Combine other tools for manual refinement.
\item Has collaboration changed your creative process? (single choice): Yes, from ideation--execution to an iterative loop; No; Yes, more fragmented/experimental.
\item Assess aesthetic value (matrix, 1--5): (A) Technical aesthetics (e.g., algorithmic randomness); (B) Prompt design is central; (C) AI style homogenizes; (D) Collaboration expands boundaries.
\item Impact on arts learning (multiple choice): (A) Tool operation skills; (B) Reflection on traditional artistic languages; (C) Critical filtering; (D) Cross-media thinking; (E) Deeper reflection on authorship/ethics.
\item Open-ended: Most profound co-creation experience.
\item Open-ended: Can AI collaboration be a new mode of arts education? Why?
\end{enumerate}

\subsection*{B. Selected descriptive results (expanded)}
\noindent The main text reports Tables~\ref{tab:profile}--\ref{tab:benefits}. For completeness, we additionally note these item-level distributions derived from the same dataset (N=112):
\begin{itemize}[leftmargin=*]
\item Aesthetic decision-making (Q5, multiple choice): (A) 47.32\%; (B) 66.96\%; (C) 57.14\%; (D) 26.79\%; (E) 0\%.
\item Role perception details (Q3, 1--5 distributions) are available on request; summary statistics are reported in Table~\ref{tab:role}.
\item Conflict-handling (Q6, ranking): top-1 choices---Accept AI output (37.5\%); Revise prompts (37.5\%); Fuse results (5.36\%); Abandon AI (18.75\%); Combine other tools (0.89\%).
\end{itemize}

\bibliographystyle{plain}
\bibliography{references}

\end{document}